\documentclass[12pt,
reprint,
superscriptaddress,
amsmath,amssymb,
aps,
prl
]{revtex4-2}

\usepackage{graphicx}
\usepackage{dcolumn}
\usepackage{bm}
\usepackage[T1]{fontenc}
\usepackage{textcomp}
\usepackage{subfigure}
\usepackage{upgreek}

\begin{document}
	
	\preprint{APS/123-QED}
	\title{Magnon-Phonon Interaction Induced Electromagnetic Wave Radiation in the Strong Coupling Region}
	
	\author{Yahui Ji}
	\affiliation
	{School of Integrated Circuits and Beijing National Research Center for Information Science and Technology (BNRist), Tsinghua University, Beijing 100084, China}
	
	\author{Chenye Zhang}
	\affiliation
	{School of Integrated Circuits and Beijing National Research Center for Information Science and Technology (BNRist), Tsinghua University, Beijing 100084, China}
	
	
	\author{Tianxiang Nan}
	\email{nantianxiang@mail.tsinghua.edu.cn}
	\affiliation
	{School of Integrated Circuits and Beijing National Research Center for Information Science and Technology (BNRist), Tsinghua University, Beijing 100084, China}



\begin{abstract}
We theoretically study the electromagnetic wave radiation of magnons driven by acoustic phonons in systems with strong magnon-phonon interaction. We evaluate the field dependence of radiation intensity spectra which exhibits the avoided crossing, a characteristic of strongly coupled systems. At the crossover where the magnon and phonon eigenstates are hybridized, we demonstrate the existence of two resonant radiation frequencies with circular polarization and the enhancement of antenna radiation efficiency by over 100 times. 
Our results open up possibilities of developing ultra-compact antennas by using the hybridized magnon-phonon mode.

\end{abstract}
\maketitle

Magnons are the quanta of spin waves, the collective excitation of magnetically ordered materials \cite{kruglyak2010magnonics,serga2010yig}, which can be used for low-dissipation information processing and communications without moving charges \cite{chumak2014magnon,chumak2015magnon,csaba2017perspectives,mahmoud2020introduction,pirro2021advances}. 
The efficient excitation of magnons can be realized in coupling with phonons, the quanta of lattice vibrations, which allows the hybridization of magnon and phonon modes in the strong coupling region, referred as a magnon-polaron \cite{spencer1958magnetoacoustic,pomerantz1961excitation,kobayashi1973ferromagnetoelastic,kobayashi1973ferromagnetoelastic2,belyaeva1992magnetoacoustics,streib2019magnon,li2020hybrid,li2021advances}. 
Ferromagnetic magnons with their intrinsic magnetization dynamics in the GHz range can couple to mechanical vibrations at the same frequency generated from acoustic transducers, which has attracted growing attention in microwave devices \cite{weiler2011elastically,kovalenko2013new,zhang2020unidirectional}. 
Here we propose a magnon-polaron antenna that can radiate electromagnetic (EM) waves by magnetization precession driven by mechanical vibrations when magnetic and acoustic resonances are synchronized.
Because such antennas are driven by acoustic resonances, their size is comparable to the acoustic wavelength and no longer limited by the EM wavelength, leading to a massive reduction of device footprint.

This magnon-polaron antenna is in stark contrast to the recently demonstrated magnetoelectric (ME) antennas \cite{domann2017strain,zaeimbashi2019nanoneurorfid,hassanien2020theoretical,chen2020ultra,nasrollahpour2021magnetoelectric},
since the EM radiation from the ME antennas is far from the magnetic resonance condition, which is evidenced by the insensitivity of radiation power to external magnetic fields \cite{nan2017acoustically,zaeimbashi2021ultra}.
In analytical models of ME antennas, although the Landau-Lifshitz-Gilbert (LLG) equation that governs the micromagnetic dynamics has been considered \cite{yao2015bulk2,yao2018multiscale,yao2019modeling}, the theory of EM wave radiation induced by the strong magnon-phonon coupling is elusive.
It is expected that the hybridization of magnon and phonon resonances can significantly boost the radiation power, however, the analytical model of EM wave radiation in the strong coupling region has not been established.

In this Letter, we theoretically investigate the EM wave radiation of magnon resonances in a low damping ferrimagnetic insulator driven by a bulk acoustic wave (BAW) using a one-dimensional (1D) multiscale finite-difference time-domain (FDTD) model.
We demonstrate the frequency splitting of antenna radiation peaks when the magnetic resonance is brought into the acoustic resonance by tuning the external magnetic field. 
The field dependence of radiation intensity spectra exhibits the avoided crossing which reveals the magnon and phonon dispersion relation at the crossover, a direct evidence of strongly coupled systems. 
At the magnon-polaron mode, the antenna radiation efficiency can be enhanced by over 100 times compared to that outside the strong coupling region. 
In addition, we find that the antenna polarization mode can be reconfigured from a circular to an elliptical polarization depending on the strength of applied out-of-plane magnetic fields. 

\begin{figure*}[!t]
	\centering	
	\begin{minipage}[b]{0.32\textwidth}
		\subfigure[]{
			\includegraphics[width=\textwidth]{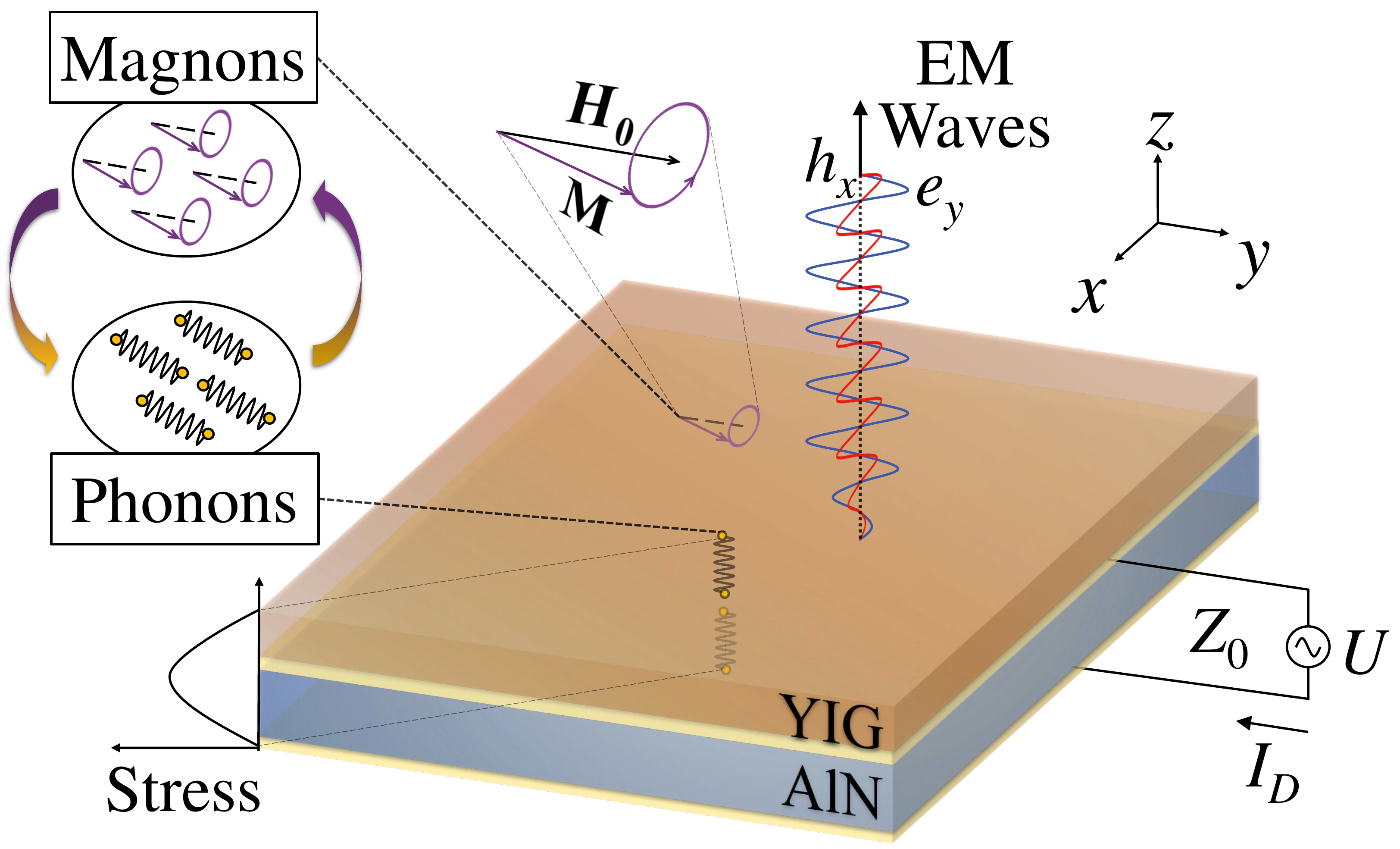}\label{FIG1a}}
		\subfigure[]{
			\includegraphics[width=\textwidth]{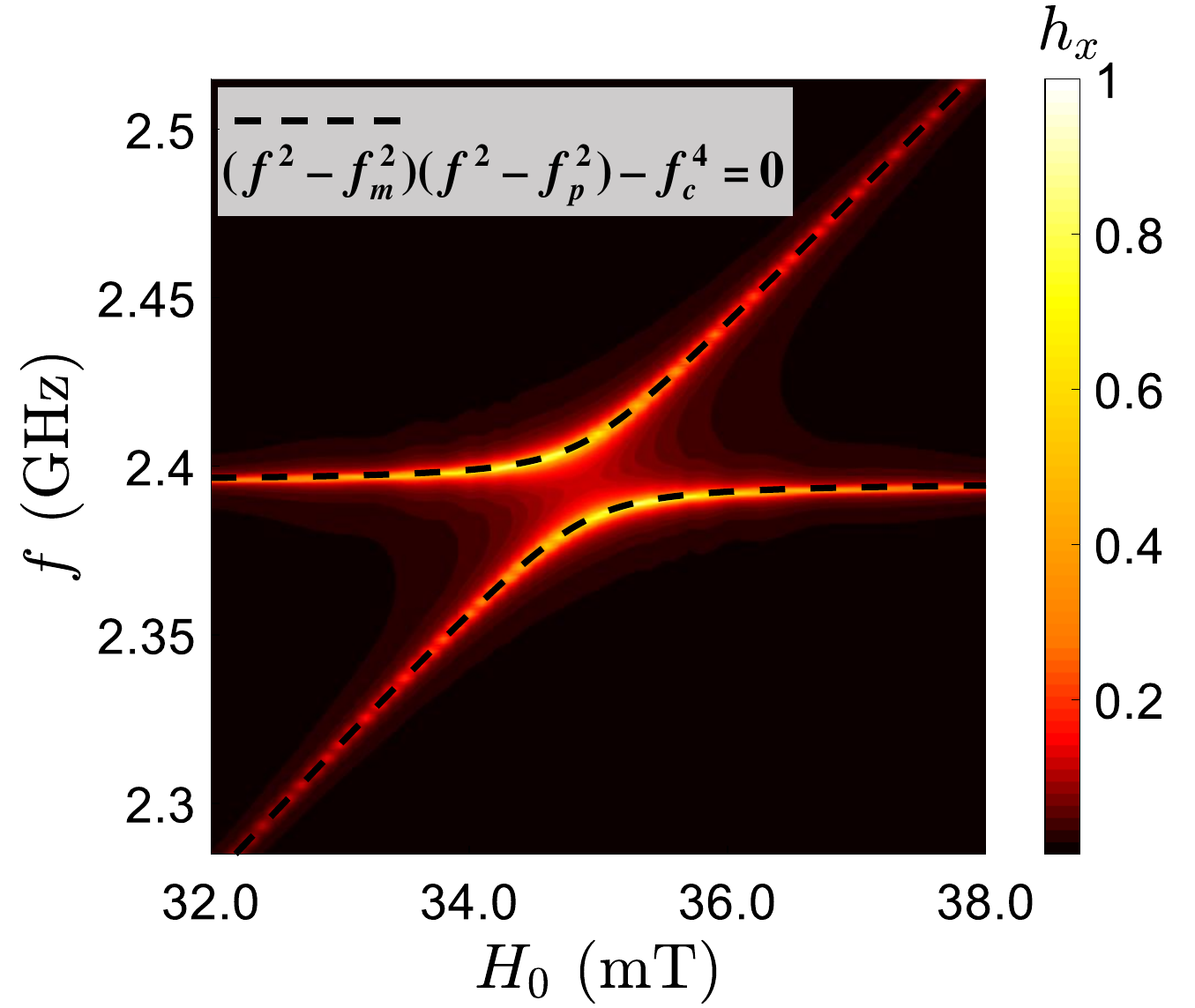}\label{FIG1b}}
	\end{minipage}
	\begin{minipage}[b]{0.32\textwidth}
		\subfigure[]{
			\includegraphics[width=\textwidth]{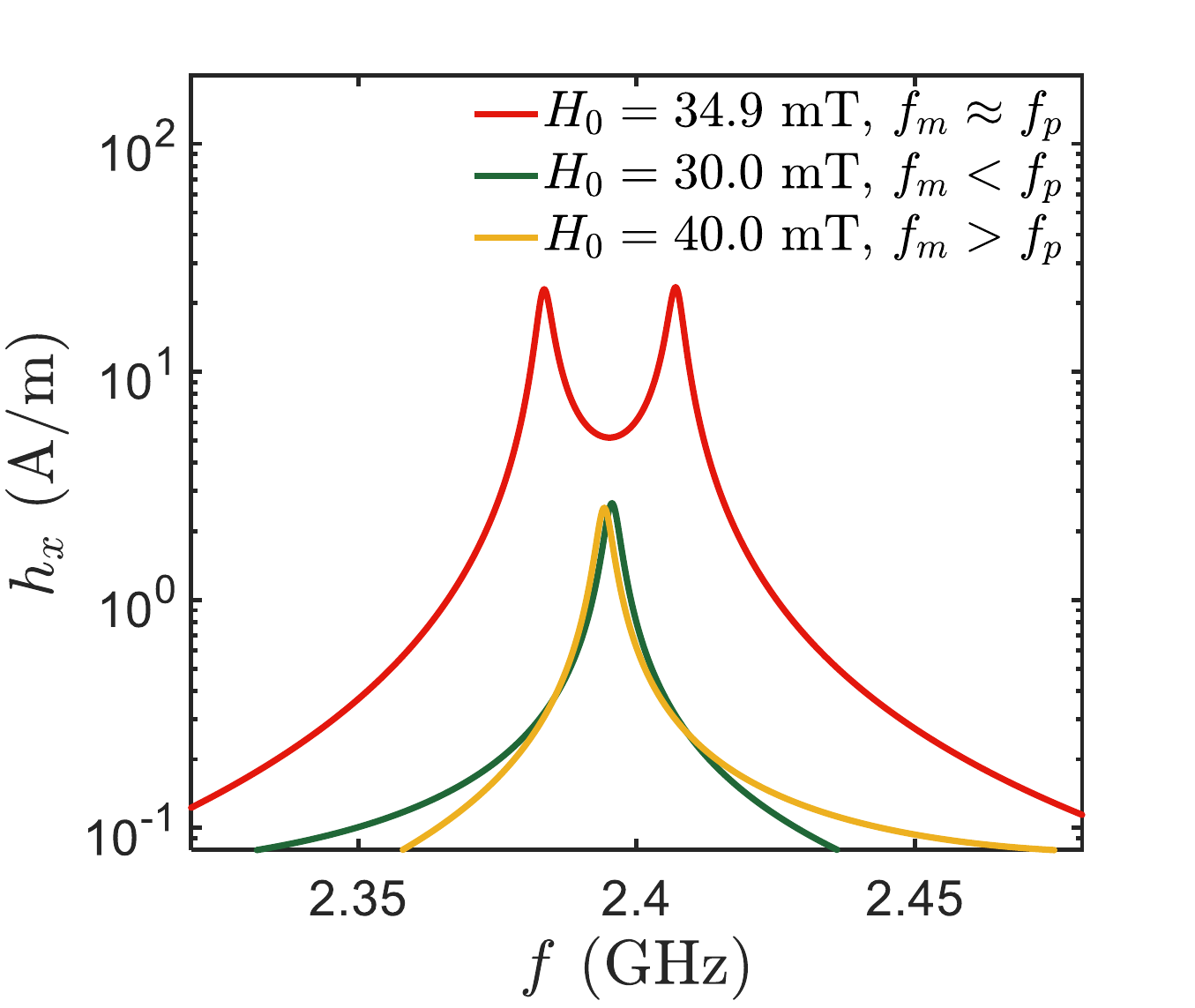}\label{FIG1c}}
		\subfigure[]{
			\includegraphics[width=\textwidth]{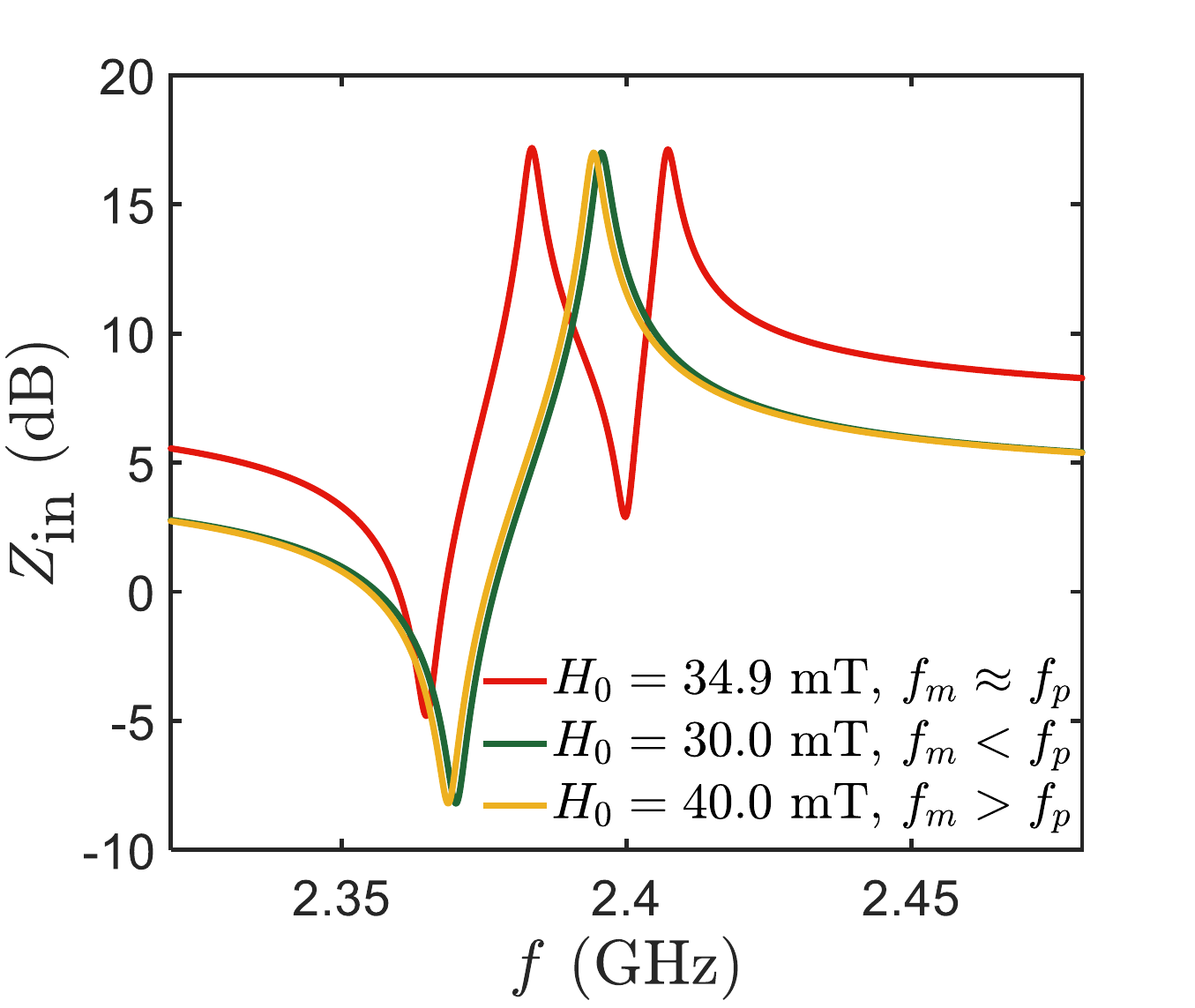}\label{FIG1d}}
	\end{minipage}
	\begin{minipage}[b]{0.32\textwidth}
		\subfigure[]{
			\includegraphics[width=\textwidth]{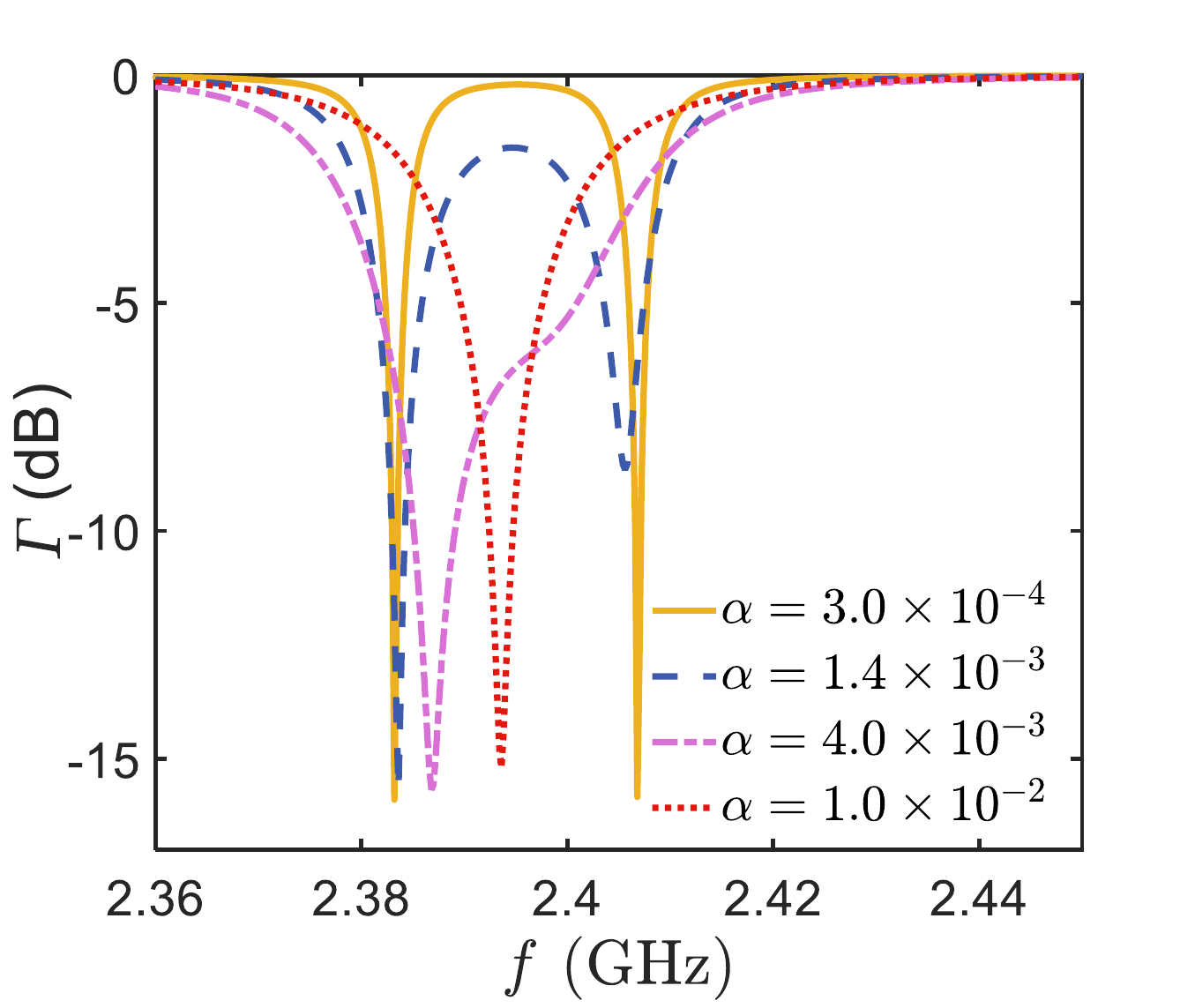}\label{FIG1e}}
		\subfigure[]{
			\includegraphics[width=\textwidth]{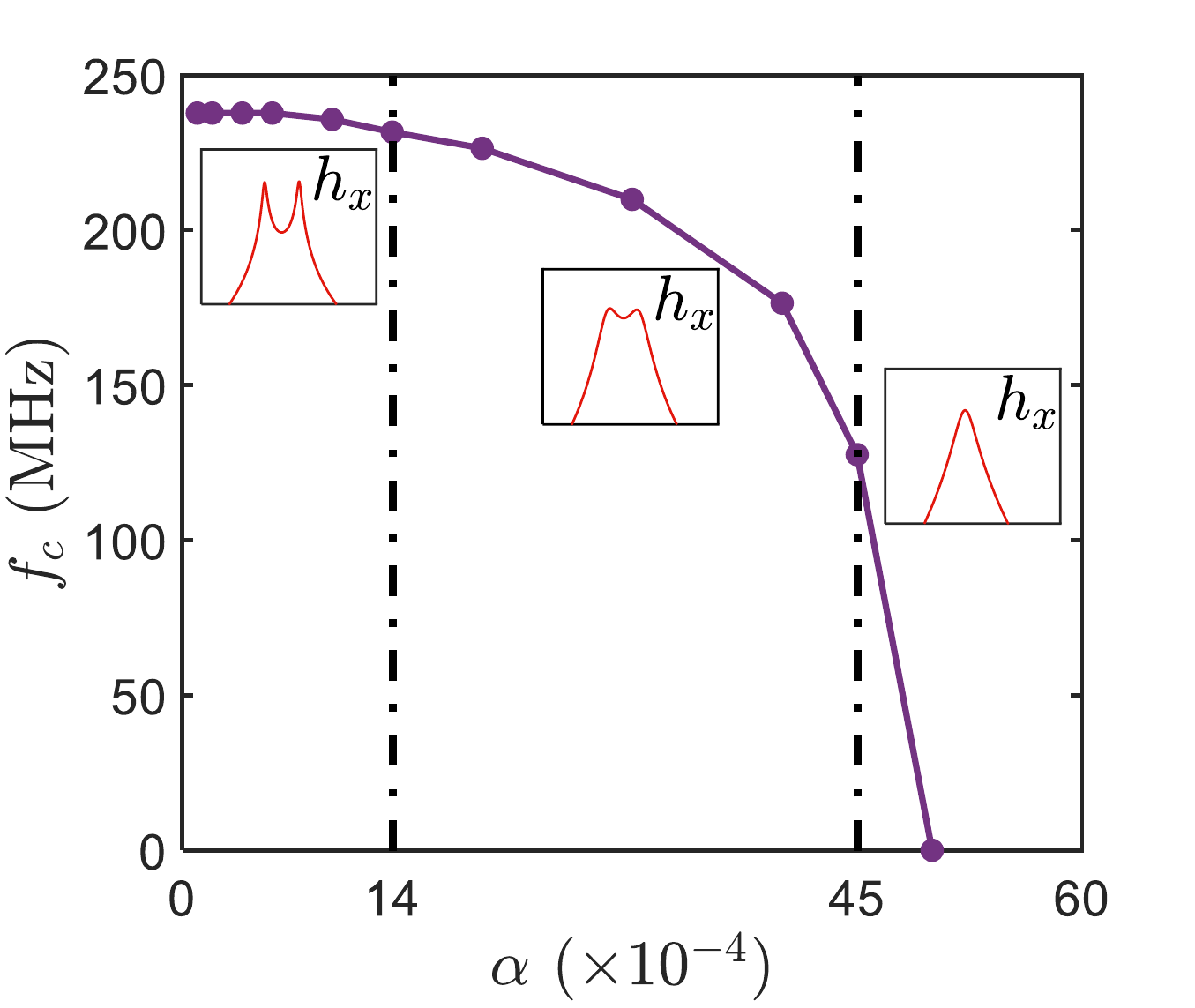}\label{FIG1f}}
	\end{minipage}
	\caption{(a) Schematic of the proposed magnon-polaron antenna structure and the EM radiation mechanism. (b) Normalized Fourier amplitude spectra of radiation magnetic field (along the $x$ axis) $h_x$ as a function of the magnetic bias field $H_0$. (c) Frequency spectra of $h_x$ at different magnetic bias fields. (d) Frequency spectra of the input impedance $Z_{\textrm{in}}=U/I_D$ at different magnetic bias fields. (e) Frequency spectra of the reflection coefficient $\varGamma=\left(Z_{\textrm{in}}-Z_0\right)/\left(Z_{\textrm{in}}+Z_0\right)$ for various Gilbert damping constants $\alpha$ of YIG. (f)  Magnon-phonon coupling factor $f_c$ as a function of $\alpha$. The insets are the representative radiation field spectra with different $\alpha$.}\label{FIG1}
\end{figure*}

The EM wave radiation mechanism and the magnon-polaron antenna structure are illustrated in FIG. \ref{FIG1a}. The antenna is composed of a ferrimagnetic/piezoelectric heterostructure, in which an voltage excitation $U$ or displacement current $I_D$ is applied on the piezoelectric layer to stimulate a mechanical vibration that can be immediately transferred to the upper ferrimagnetic layer. This consequently produces a standing acoustic wave in the $z$ direction of the  heterostructure. An in-plane magnetic bias field $\boldsymbol{H}_0$ is applied in the $y$ direction to modify the ferromagnetic resonance (FMR) frequency. Due to the strain mediated ME coupling, the piezoelectric driven oscillating strain induces magnetization dynamics $\boldsymbol{M}$ in the magnetostrictive layer, which is governed by the LLG equation as 
\begin{equation}
\frac{\partial \boldsymbol{M}}{\partial t}={{\mu }_{0}}\gamma \left( \boldsymbol{M}\times {{\boldsymbol{H}}_{\text{eff}}} \right)-\frac{\alpha }{\left| \boldsymbol{M} \right|}\boldsymbol{M}\times \frac{\partial \boldsymbol{M}}{\partial t}
\end{equation}
where ${\mu }_{0}$, $\gamma$ and $\alpha$ are the permeability of vacuum, the gyromagnetic ratio constant and the Gilbert damping constant, respectively.
${{\boldsymbol{H}}_{\text{eff}}}$ is the effective magnetic field which includes the magnetic bias field $\boldsymbol{H}_0$, the radiation magnetic field $\boldsymbol{H}$ and the ME coupling induced equivalent field ${{\boldsymbol{H}}_{\text{ME}}}$.
When the magnon and phonon resonances are hybridized, the magnon-polarons efficiently radiate EM fields $\boldsymbol{E}$ and $\boldsymbol{H}$ with dynamic components $e_y$ and $h_x$ propagating into free space, which is described by Maxwell's equations as
\begin{equation}\label{Eq16}
\nabla \times \boldsymbol{H}={{\varepsilon }_{r}}{{\varepsilon }_{0}}\frac{\partial \boldsymbol{E}}{\partial t}
\end{equation}
\begin{equation}\label{Eq15}
\nabla \times \boldsymbol{E}=-{{\mu }_{0}}\frac{\partial(\boldsymbol{H}+\boldsymbol{M})}{\partial t}
\end{equation}
where ${\epsilon }_{0}$ and ${\epsilon }_{r}$ are the permittivity of vacuum and the relative dielectric constant, respectively.
In our FDTD model, the LLG equation and Maxwell's equations are coupled with Newton's equations that govern the dynamics of acoustic waves.
The detailed description of the FDTD model can be found in Supplementary Information \cite{ji2022acoustically}.

Here we choose a model system of aluminum nitride (AlN)/yttrium iron garnet (YIG) heterostructure, because the AlN thin film is the most widely used  piezoelectric material for BAW resonators \cite{bhugra2017piezoelectric} and YIG is the bench-marked ferrimagnetic insulator with ultra-low damping \cite{spencer1959low,cherepanov1993saga,chang2014nanometer}.
In our 1 $\upmu$m AlN/750 nm YIG heterostructure, the standing phonon mode or the mechanical resonance frequency $f_p$ is governed by the thickness of AlN and the sound velocity of YIG and AlN, which is simulated to be $2.39$ GHz (See Supplementary Information \cite{ji2022acoustically} for the material parameters and the simulation settings). 
The magnon mode or the FMR frequency $f_m$ of YIG can be shifted by the in-plane magnetic bias $H_0$ following the Kittel equation $f_m=\frac{\mu_0\gamma}{2\pi}\sqrt{H_0(H_0+M_s)}$, where $M_s$ is the saturation magnetization intensity.  
FIG. \ref{FIG1b} shows the calculated radiation field $h_x$ as functions of magnetic bias field $H_0$ and frequency, which exhibits the avoided crossing, indicating the hybridization of magnon and phonon modes in the strong coupling region and the formation of magnon-polarons. 
The numerically calculated field dependence of radiation spectra can be fitted to a characteristic equation for strongly coupled systems $(f^2-f_m^2)(f^2-f_p^2)-f_c^4=0$, where $f_c$ is the coupling factor \cite{dreher2012surface,berk2019strongly}.
The coupling factor that determines the strength of the magnon-phonon coupling is directly associated to the magnetostriction coefficient, the magnetic moment and wave vectors of magnons and phonons \cite{an2020coherent}.

To evaluate the radiation properties of the magnon-polaron antenna, we compare the frequency spectra of the radiation field [FIG. \ref{FIG1c}] and the input impedance $Z_{\textrm{in}}=U/I_D$ [FIG. \ref{FIG1d}] at different magnetic bias fields.
When $f_m$ is away from $f_p$ ($f_m<f_p$ or $f_m>f_p$), one radiation peak is presented in FIG. \ref{FIG1c}, which is only corresponding to the mechanical resonance in FIG. \ref{FIG1d} at the same frequency. This is consistent with that has been observed in ME antennas \cite{nan2017acoustically,zaeimbashi2021ultra}.
As the magnon resonance is brought closer to the phonon resonance ($f_m\approx f_p$), the radiation peak is splitted into two peaks [FIG. \ref{FIG1c}] with their intensities enhanced by over 10 dB, which leads to possible applications of magnon-polaron antennas with the radiation field much higher than that of ME antennas.  
Interestingly, in the hybridized mode, the impedance spectra also show two mechanical resonances [FIG. \ref{FIG1d}], because the hybridized mode does not exist in a specific eigenstate, but rather has both magnonic and phononic characters.
The impedance measurement of magnon-phonon systems might therefore serve as a sensitive probe for the observation of magnon-phonon dynamics. 

The antenna resonance of the hybridized mode is sensitive to the Gilbert damping constant, since the magnon-phonon coupling strongly depends on the loss rate of the system.
To better illustrate the evolution of antenna resonance characteristics, we show the spectra of the reflection coefficient defined as $\varGamma=\left(Z_{\textrm{in}}-Z_0\right)/\left(Z_{\textrm{in}}+Z_0\right)$ for various Gilbert damping constants $\alpha$ in FIG. \ref{FIG1e}, where $Z_0$ is the characteristic impedance of 50 $\Omega$.
As the Gilbert damping constant increases from $3.0\times10^{-4}$ (typical damping constant of YIG) to $1.0\times10^{-2}$  (typical damping constant of ferromagnetic metals), two symmetric resonances in $\varGamma$ induced by the hybridized mode are gradually merged into one resonance solely induced by the phonon mode. 
FIG. \ref{FIG1f} summarizes the magnon-phonon coupling factor $f_c$ and the corresponding representative profiles of radiation field spectra as a function of Gilbert damping constants $\alpha$.
We find that the magnon-phonon coupling factor $f_c$ drastically decreases when $\alpha$ exceeds $1.4\times10^{-3}$, which implies a threshold-like behavior.

\begin{figure}[htbp]
	\centering
	\includegraphics[width=0.44\textwidth]{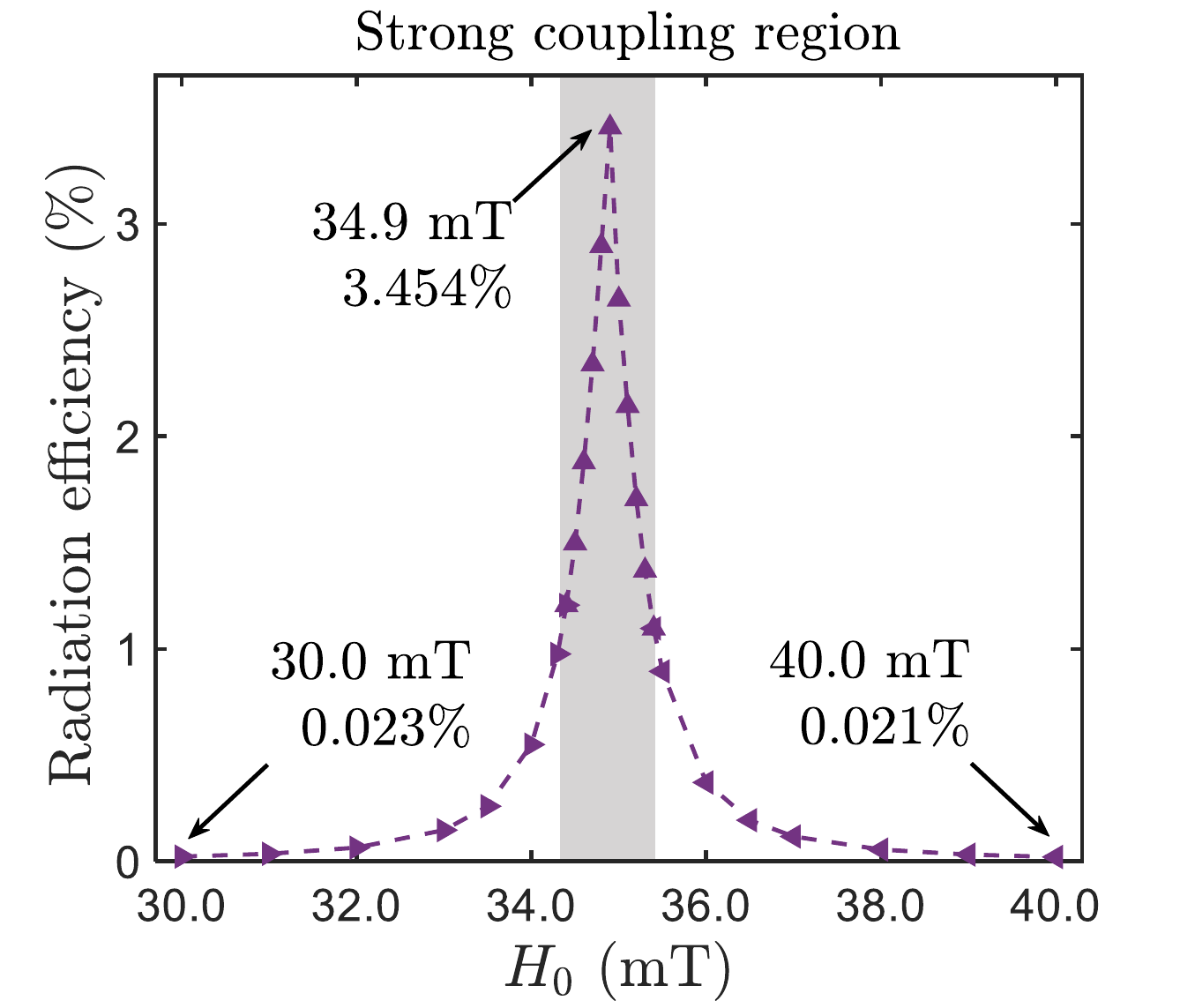}
	\caption{Antenna radiation efficiency as a function of magnetic bias field $H_0$.}\label{FIG2}
\end{figure}

We further calculate the antenna radiation efficiency (the ratio of radiated power to input power) with and without the hybridization mode by taking proper loss mechanisms into account.
For conventional antennas, the radiation efficiency is determined by the ohmic loss, while for the magnon-polaron antenna, the radiation efficiency is dominated by the mechanical viscosity and the magnetic damping.
We consider the mechanical quality factor of the acoustic structure to be around $600$ and the Gilbert damping constant of YIG to be $3.0\times10^{-4}$ in numerical simulation. 
FIG. \ref{FIG2} shows the antenna radiation efficiency as a function of magnetic bias field $H_0$.
Beyond the strong coupling region, the radiation efficiency is in the order of $0.02\%$.
While in the the strong coupling region, the radiation efficiency can be enhanced by over two orders of magnitude with a maximum value around $3.5\%$. 
We note that the antenna efficiency can be further enhanced by engineering the low loss magnetic materials with higher magnetostriction constant.
Magnesium aluminum ferrite (MAFO) thin film, for example, has been recently demonstrated with a high magnetostriction constant while maintaining a small Gilbert damping constant comparable to YIG \cite{li2022anisotropic}.
On the other hand, metallic ferromagnetic thin-film materials with generally large magnetostriction constant and low damping constant comparable to YIG such as CoFe alloys can also be used to enhance the radiation efficiency \cite{schoen2016ultra}.

\begin{figure}[htbp]
	\centering	
	\begin{minipage}[b]{0.23\textwidth}
		\subfigure[]{
			\includegraphics[width=\textwidth]{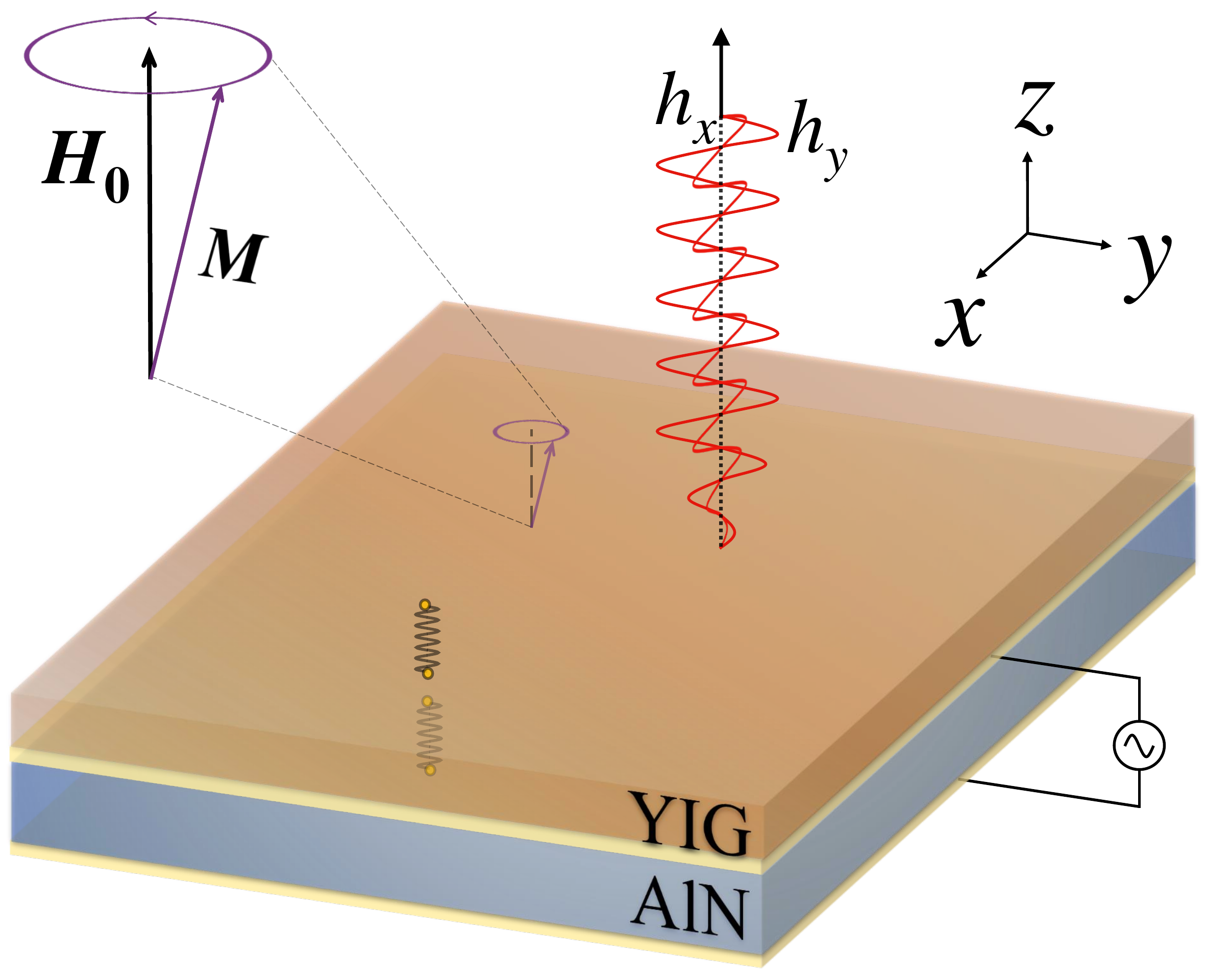}\label{FIG3a}}
		\subfigure[]{
			\includegraphics[width=\textwidth]{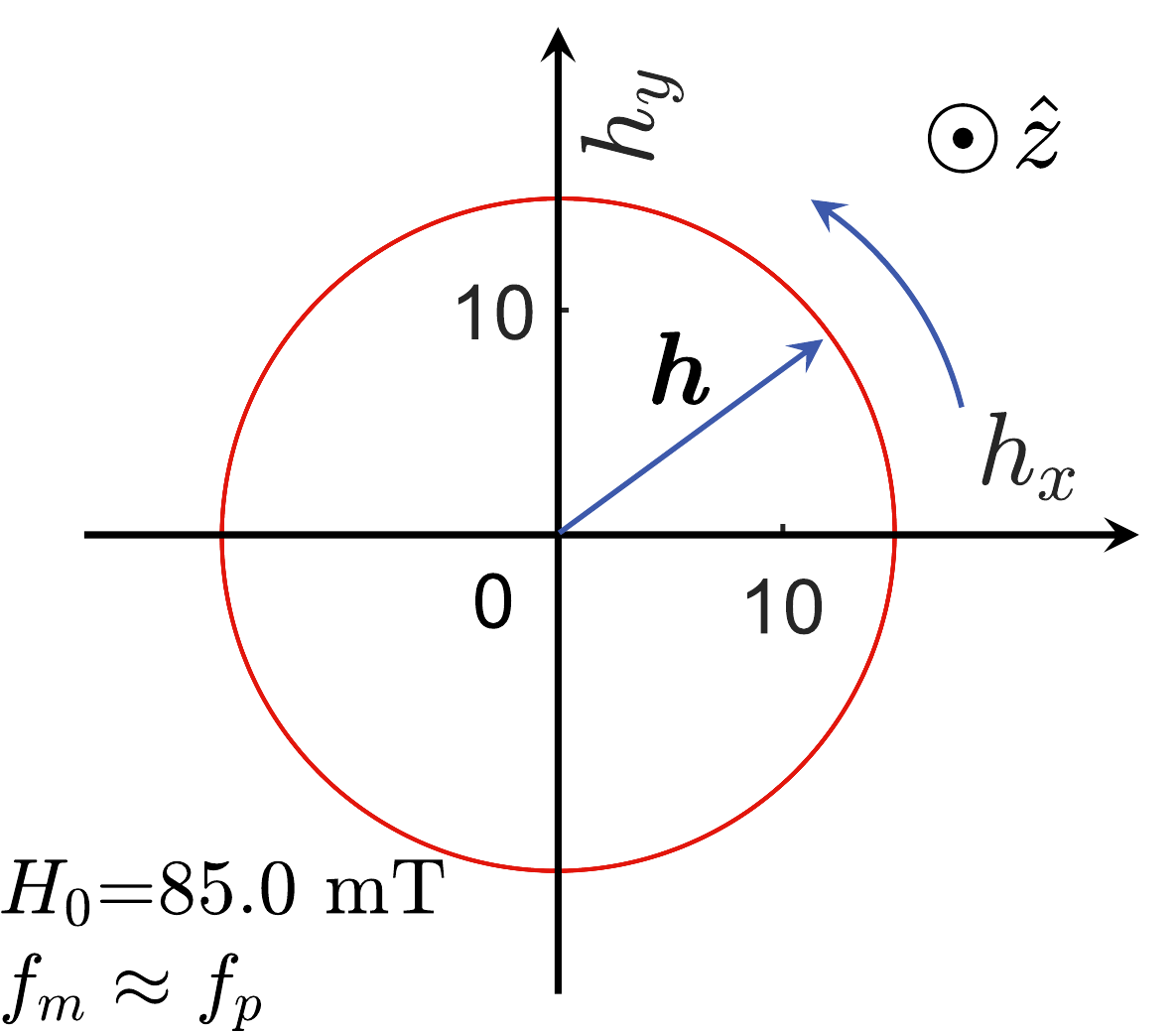}\label{FIG3b}}
	\end{minipage}
	\begin{minipage}[b]{0.23\textwidth}
		\subfigure[]{
			\includegraphics[width=\textwidth]{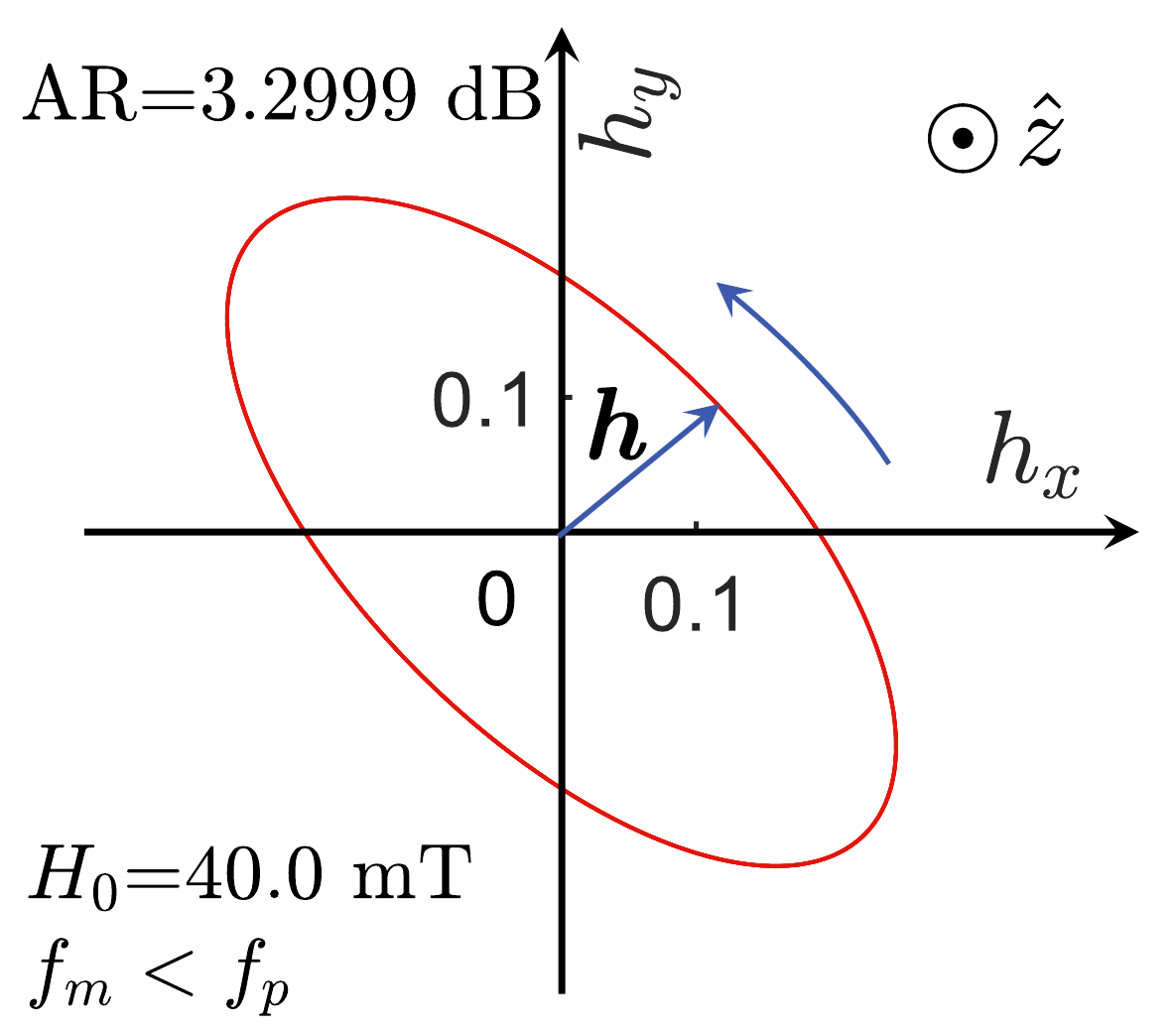}\label{FIG3c}}
		\subfigure[]{
			\includegraphics[width=\textwidth]{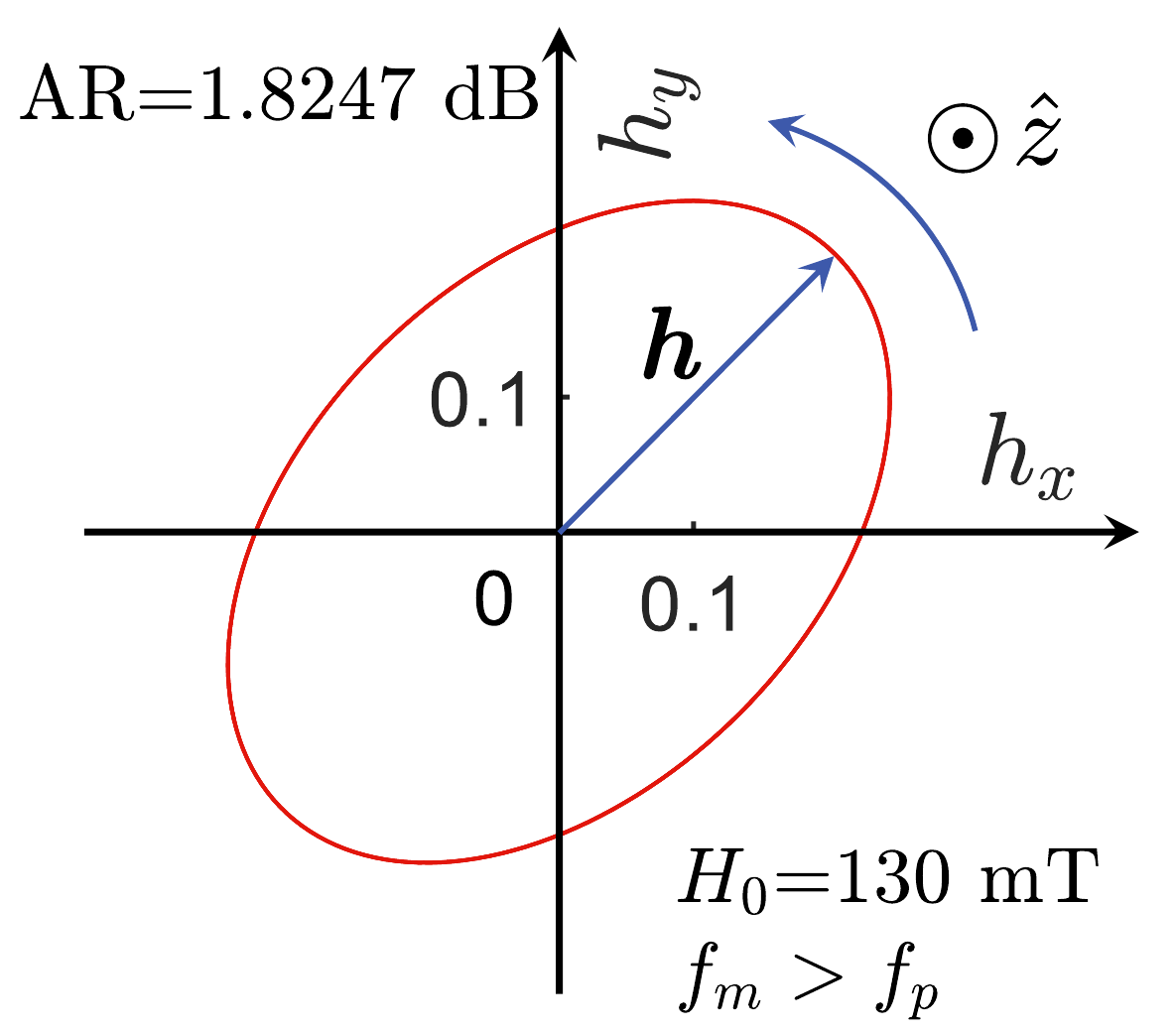}\label{FIG3d}}
	\end{minipage}
	\caption{(a) Schematic of magnon-polaron antenna under the out-of-plane magnetic bias field. Polarization modes of the antenna under different out-of-plane magnetic bias fields of (b) 85.0 mT with $f_m\approx f_p$ (c) 40.0 mT with $f_m<f_p$, and (d) 130 mT with $f_m>f_p$. The unit of $h_x$ and $h_y$ is in A/m. Note that the radiation field strength of (b) is about 100 times larger than that in (c) and (d). }\label{FIG3}
\end{figure}

To study the polarization modes of antennas, the $x$ and $y$ components of the radiation field need to be simultaneously analyzed. 
We construct an out-of-plane magnetic bias field model which allows us to calculate the radiation magnetic field $h_x$ and $h_y$ (See Supplementary Information \cite{ji2022acoustically}). 
Under the out-of-plane magnetic field bias as shown in FIG. \ref{FIG3a}, the hybridization of magnon and phonon modes takes place in a different field region since the FMR condition is different between the in-plane and out-of-plane field scenarios. 
We find the polarization modes strongly dependent on the magnetic bias field, where FIG. \ref{FIG3b} to FIG. \ref{FIG3d} show the evolution of the antenna polarization mode with increasing magnetic fields across the strong coupling region. 
In the strong coupling region ($f_m\approx f_p$), a right-hand circular polarization (RHCP) is observed. 
Such polarization mode induced by the magnon-phonon coupling is in sharp contrast to the ME antennas which have been demonstrated to have a linear polarization \cite{nan2017acoustically}. 
From the antenna application point of view, circularly polarized antennas are of special interest as EM transmitting and receiving are locally omnidirectional, which can combat signal fading from a pair of misaligned linearly polarized antennas \cite{wu2021broadband}.
Outside the strong coupling region ($f_m\neq f_p$), the magnon-polaron antenna shows a right-hand elliptical polarization (RHEP), in which the axial ratio (AR) of elliptical polarization (the ratio between radiation field amplitude of long axis and short axis) strongly depends on the out-of-plane magnetic bias field.
These behaviors can be attributed to the circular or elliptical magnetization precession under different magnetic bias fields. 
The numerical results of the polarization mode with both isotropic and anisotropic magnetostriction constant is in a good agreement with our analytical calculations (See Supplementary Information \cite{ji2022acoustically}).
The magnetic field tunable polarization mode in the magnon-polaron antenna suggests applications in reconfigurable antennas \cite{simons2002polarization,ojaroudi2020reconfigurable}.

In summary, we have theoretically demonstrated that the hybridized magnon-phonon dynamics can lead to efficient EM wave radiation in a ferrimagnetic/piezoelectric heterostructure.
In the strong magnon-phonon coupling region, we have shown that the antenna radiation efficiency can be enhanced by two orders of magnitude.
The formation of the magnon-polarons leads to the frequency splitting of antenna radiation peaks which can be tuned by engineering the magnetostriction coefficient and the Gilbert damping constant of the magnets. 
Such dual splitting resonances can be used to transmit binary information with frequency shift keying (FSK) modulation, which effectively broadens the bandwidth of the antenna. 
Our results are in sharp contrast to ME antennas in which the magnetic and acoustic resonances are significantly mismatched.
The magnon-polaron induced EM radiation opens up intriguing perspectives for developing highly efficient electrically small antennas that have the size comparable to the acoustic wavelength, and sheds lights on magnonic devices for information communications.

\begin{acknowledgments}
This work was supported by the National Key R\&D Program of China (2021YFA0716500, 2021YFB3201800), National Natural Science Foundation (52073158, 52161135103, 62131017), and the Beijing Advanced Innovation Center for Future Chip (ICFC).
\end{acknowledgments}
\bibliography{JYH2022PRL}
\end{document}